\documentclass[onecolumn]{aa} 

\usepackage{graphicx}
\usepackage{txfonts}
\graphicspath{{FIP_figures/}}
\usepackage{natbib}
\usepackage{longtable}

\begin{document}

   \title{Elemental composition in quiescent prominences }


   \author{S. Parenti
          \inst{1}
          \and
          G. Del Zanna\inst{2}
          \and
          J.-C. Vial
          \inst{1}
          }

   \institute{Institut d'Astrophysique Spatiale, 
 CNRS, Univ. Paris-Sud, Universit\'{e} Paris-Saclay\\
 Bat. 121, F-91405 Orsay, France\
              \email{susanna.parenti@ias.u-psud.fr}
         \and
             DAMTP, Centre for Mathematical Sciences, Wilberforce Road Cambridge, UK\\
             }

   \date{Received ; accepted }

 
  \abstract
  {The first ionization potential (FIP) bias is currently used to trace the propagation of solar features ejected by the wind and solar eruptions (coronal mass ejections).  The FIP bias also helps us to understand the formation of prominences, as it is a tracer for the solar origin of prominence plasma. }
   {This work aims to provide elemental composition and FIP bias in quiescent solar prominences. This is key information to link these features to remnants of solar eruptions measured in-situ within the heliosphere and to constrain the coronal or photospheric origin of prominence plasma. }
   {We used the differential emission measure technique to derive the FIP bias of two prominences. Quiet Sun chromospheric and transition region data were used to test the atomic data and lines formation processes. We used lines from low stage of ionization of \ion{Si}{}, \ion{S}{}, \ion{Fe}{}, \ion{C}{}, \ion{N}{},   \ion{O}{}, \ion{Ni}{},  \ion{Mg,}{} and \ion{Ne}{}, constraining the FIP bias in the range  $4.2 \le \log T \le 5.8$. We adopted a density-dependent ionization equilibrium. }
  {We showed that the two prominences have photospheric composition. We confirmed a photospheric composition in the quiet Sun. We also identified opacity and/or radiative excitation contributions to the line formation of a few lines regularly observed in prominences.}
  {With our results we thus provide important elements for correctly interpreting the upcoming Solar  Orbiter/SPICE spectroscopic data and to constrain prominence formation.}
   \maketitle
%

\section{Introduction} \label{intro}

One of the key open  questions about solar prominences is the process(es) of their formation. The two main ideas are the emergence of their magnetic structure from the photoshere, or reorganization of the photospheric and coronal field through reconnections \citep{parenti14, vial15}. The properties of the filament  will be different if the plasma has a photospheric origin or a coronal one. The elemental composition and the first ionization potential (FIP) bias are strongly dependent on this.
The FIP bias is present when elements with FIP lower than about 10 eV are enhanced compared to those with higher FIP, with respect to photospheric values. The amplitude of this effect varies depending on the solar region, the temperature 
and the age of the observed structure \citep[see, e.g.,][]{brooks15, laming15}.
Contradictory results in terms of FIP bias have been published in the literature, with results more 
variable in active and flaring regions \citep[see e.g.,][]{baker15}. For a full review 
on the subject, see \cite{laming15} and \cite{delzanna18}.

Knowing the plasma composition is now particularly relevant for the upcoming ESA-NASA Solar Orbiter mission  \citep{muller13}, as the  
FIP  bias is considered to be a good tracer for mapping the origin of the solar wind as observed in-situ. 
While other plasma characteristics (e.g., temperatures, ionization state) can easily be modified by several processes in the heliosphere,
the chemical composition is less affected, once the fractionation processes have taken place in the solar chromosphere and 
the plasma has been released into open field lines. For this reason, it is important to make a direct comparison 
between  the FIP bias as measured, for example, in the outer corona \citep[e.g.,][]{raymond97,parenti00}  and in the solar wind. 
 The Solar Orbiter suite carries, among others, the UV spectrometer {\it Spectral Imaging of the Coronal Environment} (SPICE, \cite{cadwell17}) and in-situ {\it Solar Wind Analyser} (SWA, \cite{muller13}) instruments, which are designed for this purpose. SWA is composed of three sensors: {\it Electron Analyzer System} (EAS),  {\it Proton-Alpha Sensor} (PAS) and {\it Heavy Ion Sensor} (HIS). In particular, HIS will provide, in addition to other properties, the charge state of all the ions.

Solar prominences may erupt and can be accompanied by a coronal mass ejection (CME) \cite{webb12}. 
Observations of erupting prominences in the UV corona show that prominences cool material can expand 
and be heated \citep[e.g.,][]{heinzel16, lee17}. 
Several plasma properties have been deduced including, in a few cases, composition and FIP bias \citep{ciaravella00}.
However, as the CME propagates, the emission of the structure fades away and the identification of the prominence plasma  
within the structure is a difficult task. 
 Exemple of UV and white light CMEs are found, for example, in \cite{webb12, giordano13, gopaslwamy15}. 

 Identification of prominence material within an interplanetary coronal mass ejection 
\citep[ICMEs, see, e.g.,][]{zurbuchen16, kilpua17} has been done using in-situ data. 
This identification  generally relies on the presence of low ionization stages of the plasma in the internal and 
much denser part of the structure \citep{burlaga98,lepri10, song17}. 
 
Nevertheless, we highlight the fact that during a CME, it is not only the prominence material is ejected, 
but also the chromosphere below and the surrounding corona participate in the eruption. 
The erupting area is also affected by reconnection processes  which heat and mix the plasma of  different structures. 
These processes  may change the composition and FIP bias. 
In order to confirm the presence of remnants of prominence material within the ICME,  
there is thus the need for a reliable  reference value for prominence composition, 
and possibly its  evolution during an eruption. 

To measure the elemental composition in prominences and  filaments is quite a complex problem, as most of the core, dense plasma 
is optically thick and not in LTE. The combination of radiative transfer modeling,   multilevel atom, and observations is needed to infer the plasma properties. 
For this reason,  composition measurements of filament cores are limited to the lighter elements. 
However, the plasma becomes optically thin in the prominence-corona-transition region (PCTR, \cite{parenti15}), 
so more direct diagnostic techniques may be applied in prominences that are observed near the limb. 
The very few measurements in quiescent prominences reported  \citep[see, e.g.,][]{spicer98, baker18} suggest a photospheric composition. 
Unfortunately, these UV measurements are often limited to the analysis of a few ions and/or charge states, 
while broadening  the analysis over a wider range of temperatures and elements would provide a  clearer picture. 
An earlier analysis was carried out \citep{parenti07} (hereafter PAR07) using spectral atlas data \citep{parenti04, parenti05a}. 

The goal of the present paper is to infer the  composition of quiescent prominences 
by mapping  as many elements as possible for the low and high FIP groups, to improve the future comparisons with in-situ data. We also provide a key constraint to the prominence formation process.
In this paper we revise the PAR07 and \cite{gunar11} (hereafter GUN11) results on prominence composition by: 
a) extending the analysis to other lines and ions, in particular by adding low FIP ions (\ion{Fe}{iii}, \ion{Ni}{ii}); 
b) revising the atomic physics calculations by adding density effects  to the ionization fraction;
c) testing for opacity effects and other atomic processes that may affect the spectral lines formation;
d) providing a direct comparison with quiet Sun on-disk observations using the same set of atomic data and lines.  


 \section{Observations and previous results}\label{data}

The data used for this analysis are from the SOHO/SUMER instrument \citep{wilhelm95} which operated in the UV  780–1610 \AA~ (first order) and 390–805 \AA~ (second order). We present results for two  prominences and quiet Sun taken as reference spectra. The first prominence (a quiescent one), which we will call PRM1, was observed on 8 October 1999 during a MEDOC campaign. 
The  $0.3^{\arcsec} \times 120^{\arcsec}$ slit was positioned at the feature center and a sequence of exposures was made to obtain spectra from about 800 Å to 1600 Å (full details of the observation program can be found in \cite{parenti04}, e.g., Figures 1 and 5). These observations provided data with a spatial resolution of about $1\arcsec$ and a spectral resolution from about $45~ \mathrm{m\AA/pixel}$ (at 800 \AA) to about $41~ \mathrm{m\AA/pixel}$ (at 1600 \AA). {During the same observing campaign we pointed also at the QS, running the same observation program but using a larger slit ($1\arcsec$).}
These data were used to build a complete prominence (and QS) spectral atlas in the range 800--1250 \AA~ \citep[][to provide identification, total intensity, line peak and width]{parenti04, parenti05a}, and derive several plasma properties: central temperature \citep{parenti05b}, differential emission measure, non-thermal velocities \citep{parenti07}. The data processing, which includes instrument corrections, as well as radiometric and wavelength calibrations, are those available in $SolarSoft$ and are detailed in \cite{parenti04}. 

The second prominence, called PRM04, was observed using a similar observing program during a MEDOC campaign on the 7th and 8th June 2004. The observation was made on low-lying remnants of the prominence after a partial eruption using the $1^{\arcsec} \times 120^{\arcsec}$ slit. These data were used to derive the prominence properties by using the hydrogen Lyman lines profiles and differential emission measure analysis, combined to a 2D non-LTE prominence fine-structure modeling of the Lyman spectra \citep{gunar11}.

Among the results from the analysis of these datasets were the differential emission measure distributions between the transition-region and coronal temperatures, which we define in Sect. 3.1. These results were obtained using previous versions of the CHIANTI atomic database  (v. 4.2) and suggested photospheric composition for both prominences. 
In \cite{parenti07} and \cite{gunar11} we pointed out two aspects which affect the DEM solution at lower and higher temperatures, and thus introduce uncertainties on the determination of the DEM at such temperatures. Lines emitted at low transition region temperatures may have opacity effects  \citep{doyle80, lanzafame94, brooks00, giunta15}. The intensity of these lines will be compatible with an emission measure smaller than that predicted for optically thin lines emitted at a similar temperature, introducing an inconsistency in the DEM inversion.  For higher temperature lines, there is an additional uncertainty on the amount of emission coming from the prominence-corona transition region (PCTR) with respect to the off-limb environment. For instance,   \cite{parenti12} showed that the PCTR can be hotter than expected. 
For our datasets we could not separate the two components, introducing an additional uncertainty on the DEM inversion.
Finally, it is well known that density-dependent effects can shift the formation temperature of an ion, and thus
could in principle affect the FIP bias measurement based on a DEM analysis.   
These  aspects are part of our motivation for a new analysis.

\section{Method}\label{diagno}

\subsection{Differential emission measure}

The diagnostic method used to derive the elemental composition is the differential emission measure (DEM) applied to UV optically thin collisionally excited plasma. Under these conditions the total intensity of the observed line is given by

\begin{equation}\label{eq:I}
I_o(\lambda)= \frac{1}{4\pi}  \int_l{Ab~ G(T_e, n_e)~ n_e n_H \mathrm{d}l}
\end{equation}

\noindent where $l$ is the line of sight through the emitting plasma,  $Ab$ is  the abundance of the element with respect to hydrogen, G($T_e$, $n_e$) is the contribution ~function
which contains all the atomic physics  parameters,   $n_e$ and $T_e$ are the electron number density and temperature, $n_H$ is the hydrogen number density.

The DEM is defined here as

\begin{equation}\label{eq_dem}
  \mbox{\it DEM}(T_e) = n_e n_H \frac{\mathrm{d}l}{\mathrm{d}T_e} 
\end{equation}

so that Equation \ref{eq:I} can be written as

\begin{equation}
I_{ob}(\lambda)= \frac{1}{4\pi}  \int Ab~ G(T_e, n_e)~DEM(T_e) \mathrm{d}T_e
.\end{equation}

When a set of lines from an optically thin plasma formed at different temperatures is measured, we can derive a DEM solution. This is done by using an atomic database to calculate the synthetic spectral lines intensity ($\mathrm{I_{th}}$) and minimizing the ratio $\mathrm{I_{th}}/\mathrm{I_{ob}}$. The theoretical line intensity is calculated assuming  a tabulated abundance (in our case \cite{asplund09} photospheric), see Table \ref{table:abund}. If the composition of one element in the observed plasma is different from that assumed, all of the lines from this element will show a similar (coherent) inconsistency within the DEM inversion of the amount equivalent to the correction to be made to its tabulated value. 

We also introduce the effective temperature, which is the DEM-weighted average temperature 

\begin{equation}
T_{\rm eff} = \frac{\int DEM(T) \times T \, \mathrm{d}T}{\int DEM(T)\, \mathrm{d}T}. 
\end{equation}
 

The DEMs as function of the temperature shown in this work have been obtained with a method based on a simple chi-square minimization.
We used essentially a modified version of the  \textit{xrt\_dem\_iterative2.pro} DEM inversion routine \citep{weber04}  in order to have more flexibility
in the choice of input parameters. 
This revised routine is being made available via CHIANTI version 9 \citep{dere19}. 
The DEM is modeled assuming a spline, with a  selection of the nodes, and the program 
then uses the robust chi-square fitting routine \mbox{(\textit{mpfit.pro})}.

\begin{table}[hb]
\caption{Photospheric elemental abundances from \cite{asplund09} compatible with the results of our analysis. The values are relative to \ion{H}{} expressed in logarithmic scale.}      
\label{table:abund}      
\centering      
\setlength{\tabcolsep}{3pt}     
\begin{tabular}{c c c c c c c c c }     
\hline\hline       
 \ion{C}{}&  \ion{N}{}&  \ion{O}{} & \ion{Ne}{}&  \ion{Mg}{} &  \ion{Si}{}&  \ion{S}{} &  \ion{Fe}{} &  \ion{Ni}{}\\ 
\hline     
 8.43    &  7.83     &  8.69      &   7.93     &   7.60     & 7.51    & 7.12 & 7.50     & 6.22\\ 
 \hline    
\end{tabular}
\end{table}

\subsection{Atomic processes and data}

In our previous analyses we used the CHIANTI atomic database for calculating the 
emissivity of the lines, and assumed the standard CHIANTI ion abundances, calculated 
assuming  ionization equilibrium in  the zero density approximation. This means that 
the ion charge state distributions are calculated assuming that all ions are in their 
ground states, and balancing the collisional ionization with dielectronic
and radiative recombination rates.
However, there are several processes that depend on the electron densities 
and that can  affect the ion populations, as shown by  \cite{burgess69}.
These effects are present even at coronal densities, and become 
 significant at transition region densities and temperatures 
\citep[see, e.g., the review by][]{delzanna18}.

\begin{figure}
  \includegraphics[scale=.42]{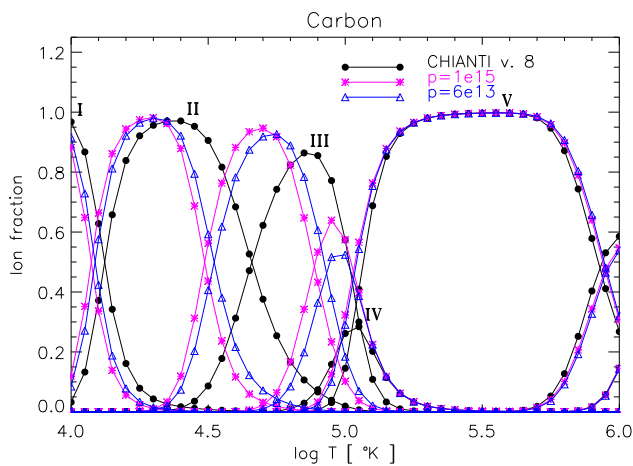}
  \caption{Carbon ionization fraction as a function of the temperature. 
Circles refer to the CHIANTI v.8 data, while  stars and triangles are obtained
from the rates in the OPEN-ADAS database, which includes  density effects. 
The assumed pressures are those used in our analysis. The pressure units are $\mathrm{cm^{-3} K}$.}
              \label{fig:ioneq}%
    \end{figure}

The inclusion of these effects requires a complex collisional-radiative modeling (CRM), as fully implemented in \cite{summers06} within the atomic data and analysis structure (ADAS)\footnote{www.adas.ac.uk} project and database, and freely available through  OPEN-ADAS\footnote{open.adas.ac.uk}. Such modeling has also recently been implemented by \cite{dufresne19}. 
For the present work we have used the results of the CRM 
available via OPEN-ADAS in the form of 
effective ionisation and recombination rates, calculated for a grid of densities
and temperatures. We have calculated the ion balances of the main elements using these
rates, assuming constant pressure.
Figure \ref{fig:ioneq} shows one example for the carbon ions: 
 the net effect is to shift the 
temperature of formation of the ions toward lower values and change the peak
ion abundances.

%
{We note that, as suggested by \cite{nussbaumer75} who studied the ionization of carbon, 
neutrals, and singly ionized ions formed at low temperatures (around 20 000 K) can be affected by 
photoionisation, depending on the density of the plasma and the photoionising flux (Dufresne $\&$ Del Zanna, 2019). }
As the main diagnostic ions used in the present analysis are mostly formed
at higher temperatures, we have neglected this process. 

Another process that we have considered instead is 
photoexcitation of a line due to the absorption of the disk radiation.
This process has a fundamental effect on the cooler lines in the outer corona, but 
for prominences  it is mainly relevant for  strong lines and relatively low densities.
To calculate this effect we have assumed that the incident disk  profile completely penetrates  the prominence plasma. We modified the CHIANTI software 
to input the photo-exciting radiation for a specific line in the level population solver.
As the process is a resonant one, the addition of photo-exciting radiation 
is taken into account by modifying the transition probability for spontanous decay. 
We note that the CHIANTI software  allows the inclusion of incident disk profile,
but this is equivalent to the method used here (and tested in \citealt{delzanna18b}), 
as long as the coronal line profile
is broader than the disk profile. We have neglected Doppler dimming effects as they 
would require velocities much higher than those typically observed in the 
quiescent prominences considered here.

For the calculations of line emissivities we used CHIANTI v.8 \citep{delzanna15}, which has significantly
improved atomic data (over previous versions) for a range of ions. 
In particular, CHIANTI v.8 introduced new data for \ion{Fe} {iii}, which we have added
to the analysis, as  lines from this ion are abundant in the SUMER spectrum. 
This is an important addition as it constrains the low transition -region temperature 
FIP bias.
We encountered several discrepancies whilst testing different ionization equilibria and comparing predicted and 
observed line radiances for this ion.
The cross-sections for collisional excitation should be reliable,
but the ion abundance is very uncertain, due to large differences in 
the  dielectronic recombination rates for the low-charge Fe ions as calculated 
by various authors. Nevertheless, as our main aim is to compare the prominence
abundances relative to the quiet Sun, any systematic differences can be accounted for,
as discussed below.

Another aspect to keep in mind is that in general atomic data  
for low charge states are difficult to calculate, and few are available. 
This is the reason why   CHIANTI  does not yet contain atomic calculations 
for all ions.  \ion{Ni} {ii} (another low FIP element) produces a few 
weak SUMER lines, but atomic data were not available in CHIANTI.

Recently, \cite{cassidy16} published a calculation of radiative rates for this ion, while previous cross-sections for collisional excitation were made available \citep{cassidy11}. However,  several discrepancies in the atomic structures of the two independent calculations have been found,
so we could only build a limited ion model to estimate the emissivity of two 
of the UV lines observed by SUMER. 
We excluded \ion{Fe}{ii} in our analysis, as the atomic data are still 
not complete. 
Aside from \ion{Ni}{} and \ion{Fe}{}, the other low-FIP element that we observed is \ion{Si}{}.
Finally, only in one dataset  could we add a \ion{Mg}{v} line to constrain the FIP bias at higher temperatures.

Sulfur is a mid-FIP element producing several lines in prominences.
 We note that \ion{S}{}  often shows variations consistently with the 
high-FIP elements, as measured in remote-sensing observations of active regions 
\citep[see, e.g., the review by][]{delzanna18}.
There are many high-FIP elements observed by SUMER.  Nitrogen is 
particularly useful because four ionization stages are observed. Their temperature formation covers almost the whole of the range we are considering, allowing us to map any possible \ion{N}{} FIP bias change within it. 

\subsection{Selection of ions and their spectral lines}

In order to identify possible deviations from our assumption (Eq. \ref{eq:I}), that is, problems in the atomic data and observations, we used the strategy of first testing the atomic data using  QS data, as its chemical  composition  is known to be close 
to photospheric  \citep[e.g.,][]{lanzafame05, parenti07}. This test provided us constraints in the choice of
 spectral lines. 
 Once this selection was made, an (almost) similar line list was also used on the prominence data. 
 Unlike  PAR07, for which the goal was to provide an overall view, 
here the main list is made of a limited number of lines from each given ion (Table \ref{table:lines}). 
During the selection process we considered several lines not included in  PAR07, 
and tested  possible temporal variations of the thermal properties of the structure.
As pointed out in PAR07, SUMER records about 40 \AA\  at a time so
 there was a time lag of about 2h between the shortest and longest wavelengths observed. 
We list in Table \ref{table:lines2} the additional lines that we have considered. 
The list includes, among others,
the \ion{C}{ii} multiplet (4 lines) at around 1323.9~\AA\ and the doublet at around 1335~\AA. 
These are slightly density-sensitive and strongly temperature-sensitive lines. 
We discuss their behavior within our datasets in the following sections.
As a general choice, we adopted lines above the hydrogen Lyman continuum (912 \AA) to avoid absorption effects. 
This implied  excluding \ion{O}{ii} lines
 and limited the choice of \ion{O}{iii} lines. We discuss this further in Sect. \ref{sec:opacity}.

Table\ref{table:lines} lists the selected lines and their intensity, together with the temperature 
at the peak  of their emissivity. 
The new lines added to those used in our previous works (PAR07, GUN11) are labeled with the ''$\circ$'' symbol. 

 \begin{table}
\caption{Line list used in this work from \cite{parenti05a} and GUN04. The intensities are given in $\mathrm{erg cm^{-2} s^{-1} sr^{-1}}$. The lines indicated with "$\circ$" extend the lists published in \cite{parenti05a} and GUN04. The lines marked with "$\bullet$" will be measured by Solar Orbiter/SPICE.}  
\label{table:lines}      
\centering 
\setlength{\tabcolsep}{3pt}         
\begin{tabular}{l r l c r r r l}     
\hline\hline       
Ion & $\lambda_{th} [\AA]$ & & $\log T$ [K]& QS & PRM1& PRM04 &\\ 
\hline     
   Si II  & 1190.416 & &4.13 &14.16  & 2.60   & 2.05 &$\circ$ \\
   Si II  & 1197.3955& &4.13 &8.59   &        & 1.21 &$\circ$ \\ 
   Si II  & 1264.738 & &4.13  &      &        & 13.73 &$\circ$ \\ 
   S II   & 1253.811 & &4.25 &11.89  & 1.20 & 0.67 &$\circ$ \\
   S II   & 1250.585 & &4.25 &5.04  & 0.37  &      &\\
   C II   & 1036.337 &$\bullet$ & 4.37 &33.01  & 10.94  & 6.40 &  \\
   C II   & 1037.018 &$\bullet$ & 4.37 &40.01  & 12.92  & 8.73 & \\    
   N II   & 1085.701 & &4.38 &43.55  & 11.48  & 6.20 & \\
   N II   & 1083.990 & &4.38 & 10.84 & 2.64   & 1.32 & \\
   S III  & 1015.496 &$\bullet$ & 4.46 &0.88   & 0.18   & &       \\
   S III  & 1200.959 & &4.46 &7.98   &        & 1.53 &$\circ$  \\
   S III  & 1194.047 & &4.46 &      &        & 0.87 &$\circ$    \\ 
   Si III & 1109.943 & &4.67 &     &        &  0.87 &$\circ$    \\ 
   Si III & 1113.25 &  &4.67 &13.29  &        &  2.05 &$\circ$   \\ 
   Si III & 1206.502 & &4.67 &549.05 & 60.89  &    &\\ 
   C III  & 977.020 &$\bullet$  & 4.85 & 518.55 & 211.91 & 277.8&  \\
   C III  & 1174.933 & &4.85 & 42.76 &        &       &\\
   C III  & 1175.7111& &4.85 & 157.10 & 16.80  & 20.51 & \\ 
   Si IV  & 1128.34  & &4.87 &12.23 & 1.66   & 1.79 &$\circ$   \\
   Si IV  & 1393.76   & &4.87 &     &         & 62.53 &$\circ$   \\
   N III  & 989.799 &$\bullet$  & 4.88 & 11.45  & 2.85   & 3.95 &   \\
   N III  & 991.577 &$\bullet$  & 4.88 &22.75 & 4.89   & 7.90   &   \\
   O III  & 525.794 &$\bullet$  & 4.93 &8.90   &        & 5.50  & \\
   O III  & 703.850 &$\bullet$  & 4.93 &      &        & 20.0   & \\
   S IV   & 1062.664 && 4.99 &3.74   & 9.48   & 0.91 &$\circ$    \\
   S IV   & 1072.974 && 4.99 &7.56  &        & 1.58 &$\circ$  \\
   C IV   & 1548.187 && 5.03  &      &        & 103.55 &\\
   N IV   & 955.334 &$\bullet$  & 5.17 &0.17  & 0.05   &  &\\
   O IV   & 555.264  && 5.18  &24.75  &        &        &\\
   O IV   & 1399.776 && 5.18   &     &        & 1.32   &\\
   O IV   & 1401.163 && 5.18  &     &        & 8.21    &\\
   N V    & 1238.821 && 5.27&57.24  &        &         &\\
   N V    & 1242.804 && 5.27  &28.97  &        &       & \\
   O V    & 1218.347 && 5.37  &62.76 & 16.08  & 17.59 &\\
   Ne V   &  572.113 && 5.46  & 2.45   &  0.63  &      & \\
   Ne V   & 1145.582 && 5.46  &     &  1.04  &       &\\
   O VI   & 1031.912 &$\bullet$ & 5.48  &247.78 &        &  &         \\
   O VI   & 1037.614 &$\bullet$ & 5.48  &116.13 &        &   &     \\
 \hline                  
\end{tabular}
\end{table}
 

 \begin{table}
\caption{Lines used to extend the list in Table \ref{table:lines}. These lines also extend the lines lists published in \cite{parenti05a} and GUN04. The lines marked with "$\bullet$" will be measured by Solar Orbiter/SPICE.
The intensities are given in $\mathrm{erg/cm^2/sr/s}$.}      
\label{table:lines2}      
\centering      
\setlength{\tabcolsep}{3pt}     
\begin{tabular}{l r l c r r r }     
\hline\hline       
Ion & $\lambda_{th} [\AA]$ & & $\log T$[K]& QS & PRM1& PRM04\\ 
\hline     
   Si II  & 1304.370 & &4.13 &        &        & 2.34\\ 
   Si II  & 1309.276 & &4.13 &        &        & 3.18\\  
   Ni II  & 1317.210 & &4.15     &         &       & 0.14 \\
   C II   & 1323.862 &(4)& 4.37& 2.29 & 0.77   & 0.46\\
   C II   & 1334.577 & &4.37 & 532.1   & 178   & 112.2 \\
   C II   & 1335.663 & (2)& 4.37 & 772.9  & 229.1  & 169.8 \\
   Fe III  &1017.254 &$\bullet$ & 4.45 & 1.08   & 0.17   & 0.12\\
   Fe III &1128.740 & &4.45  &  & 0.56   &       \\  
   Si IV  & 1393.76  & &4.87 &        &        & 63.53  \\
   Si IV  & 1402.77  & &4.87 &        &        & 27.8   \\ 
   N IV   &765.147 &$\bullet$& 5.17  &        &        & 21.70  \\ 
   Mg V   & 1324.433 && 5.45     &        &        & 0.15  \\ 
 \hline                  
\end{tabular}
\end{table}

\section{Results} 
\label{sec:results}

\subsection{Quiet Sun}
\label{sec:dem_qs}

\begin{figure}
\resizebox{\hsize}{!}{\includegraphics[scale=.4]{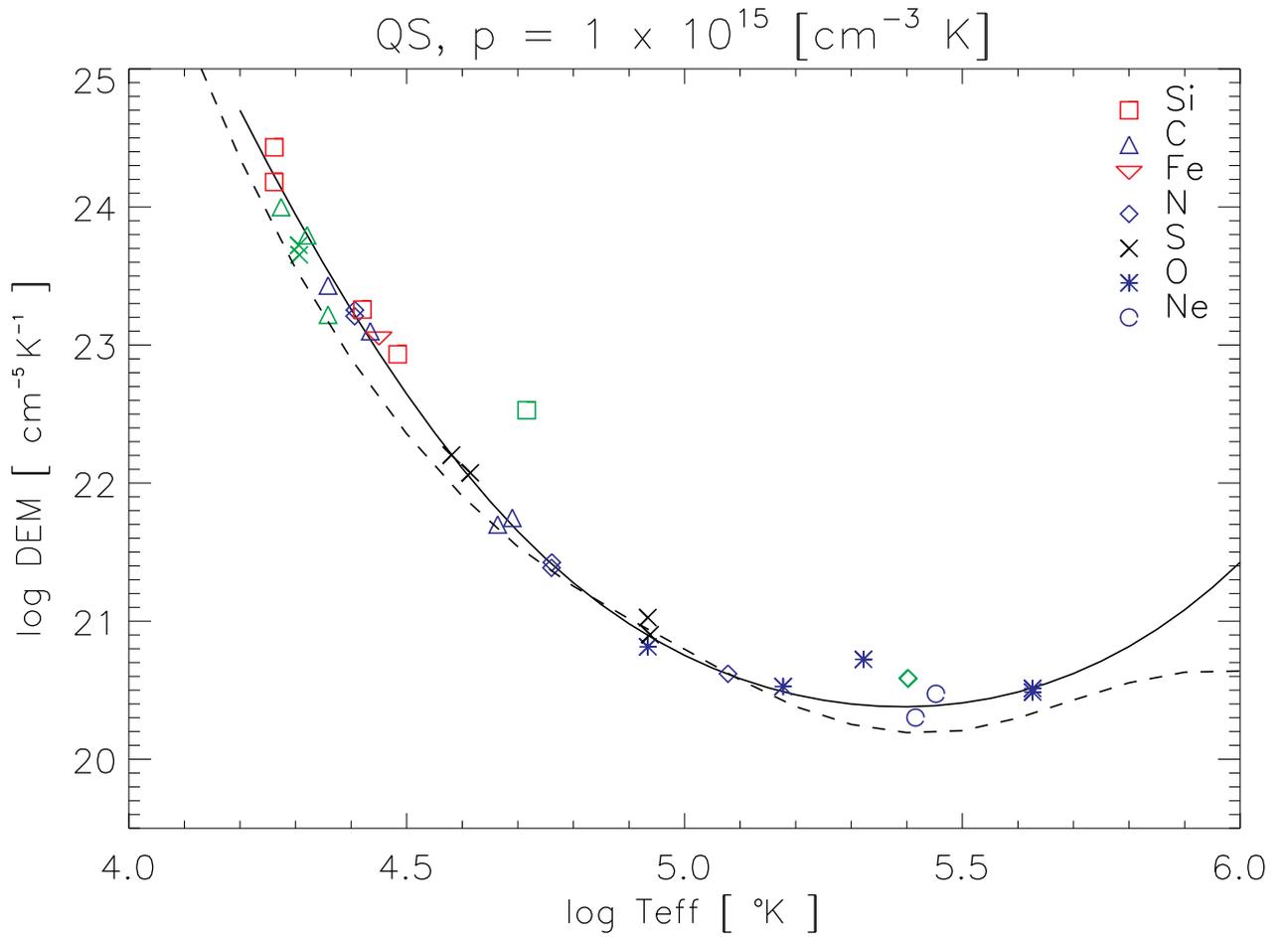}}\\
   \caption{Quiet Sun DEM. Blue represents the high FIP ions, red the low FIP ion, and black intermediate ion (sulfur). Green color is used for bright lines excluded from the DEM inversion. The dashed curve shows the DEM derived by \cite{parenti07}. The assumed photospheric abundances are from \cite{asplund09}}
              \label{fig:demqs}%
    \end{figure}

The DEM inversion was limited to the temperature range $\mathrm{4.2<\log T <5.8}$ 
to avoid the various problems previously mentioned.
 The resulting DEM as a function of the effective temperature is shown as a solid line in Figure \ref{fig:demqs}.
Details of the results are listed in Table \ref{tab:ratios}. 
In Figure \ref{fig:demqs} the two groups of high and low FIP ions are marked in, respectively, red and blue colors. 
The intermediate FIP element Sulfur is labeled in black, while green shows spectral
 lines that have been excluded from the inversion (they are indicated with "*" in Table \ref{tab:ratios}). 
For these latter lines either our DEMs produce $\mathrm{I_{th}/I_{ob}}$ above $30\%$, or the optically thin approximation does not hold. 
The behavior of these lines is discussed later. 
We note that the inclusion of the second list of lines of Table \ref{table:lines2} does not change the result. 

If we exclude the lines marked in green, this solution,
 which assumes the photosheric abundances of \cite{asplund09},  confirms the PAR07 findings. 
The  DEM  is also very close to the PAR07 result (the dashed curve in the figure). Above $\mathrm{\log T=4.5}$ we do not have suitable lines from low FIP elements and the absence of FIP bias is inferred by the intermediate FIP element sulfur. The newly included \ion{Fe}{iii} is nicely consistent with the other ions formed at a similar temperature,  adding a constraint to the FIP bias.
 
Among the lines not consistent with the resulting DEM, we have the  \ion{S}{ii} ($\log T=4.2$) lines;
their  observed intensities are too low compared to their theoretical values. We do not think this is a FIP bias effect 
(the lines from the hotter \ion{S}{} ions, \ion{S}{iii}-\ion{}{iv} are consistent with the lines from other ions formed at similar temperatures),  but it is more likely to be an opacity effect, even though the ratio of our doublet in the QS is close to the theoretical value of 2.

At low temperatures  the \ion{C}{ii} 1037.018 \AA ~~($\log T=4.3$), which is part of a resonant doublet, is too weak also: opacity may also be present here, as discussed in Sect. \ref{sec:opacity}. To provide better constraints on the DEM, we added other \ion{C}{ii} lines, which are listed in Table \ref{table:lines2}. The \ion{C}{ii} 1337--35~\AA\ doublet is often affected by opacity  (see also Sect. \ref{sec:opacity}),
so we have not used it for the DEM analysis. However, the predicted intensities are   consistent with an optically thin solution.  
The multiplet at around 1323.8 \AA~~(which is seen as a single line in our QS data) is much weaker than the boublet mentioned above,
 and we expect it to be less affected by opacity.
 The fact that its intensity is well reproduced by our DEM suggests that this case is, indeed, in the optically thin regime.

The other lines that are not consistent with the DEM are from \ion{Na}{} and \ion{Li}{} iso-electronic sequence ions (\ion{Si}{iv}, \ion{N}{v}). 
It is well known that these anomalous ions produce spectral lines with observed intensities that are typically much larger than those
predicted using the emission measures obtained from ions of other sequences
\citep[see, e.g., for a review][]{delzanna02, delzanna18}. 
The density effects included in the present modeling  improve the predicted intensities 
compared to those obtained with  the zero-density ionization equilibrium, but other effects might be at play.
Among the anomalous ions, only  \ion{O}{vi} was included in our calculation to give a constraint 
at upper temperatures.

\subsection{Prominences}
\label{sec:dem_prm}

\begin{figure*}[ht]
 \resizebox{\hsize}{!}{\includegraphics[scale=0.25]{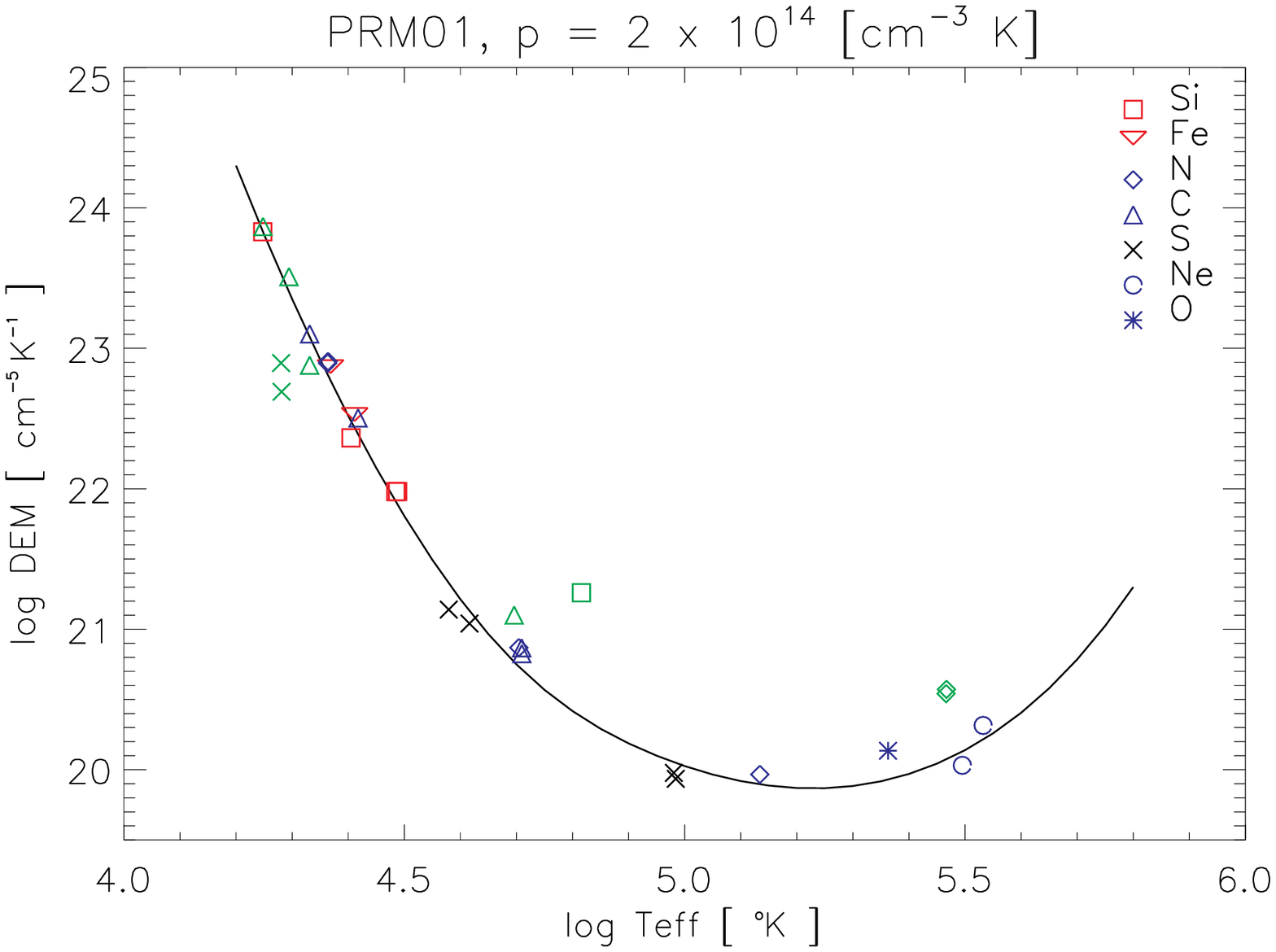}
 \includegraphics[scale=.25]{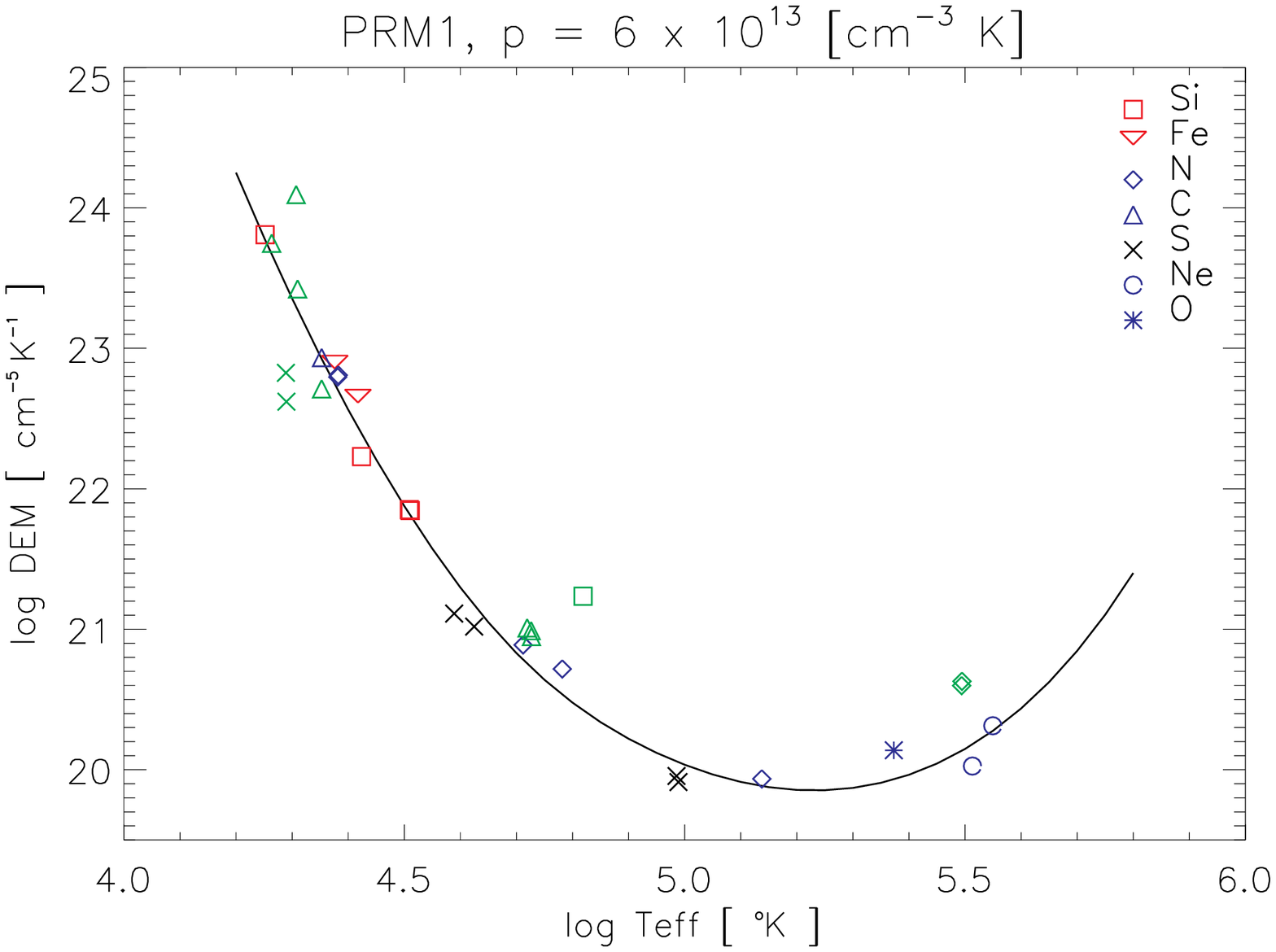}}
    \caption{PRM1 DEM assuming a pressure equal to $\mathrm{2 \times 10^{14} ~cm^{-3}K}$ (left) and $\mathrm{6\times 10^{13} ~cm^{-3}K}$ (right).  Color codes are as in Figure \ref{fig:demqs}.}\label{fig:dems_prm1}
    \end{figure*}

Figure \ref{fig:dems_prm1} shows the resulting DEMs as function of $\mathrm{T_{eff}}$ (solid line) for PRM1 assuming two different pressures: $\mathrm{2 \times 10^{14} ~cm^{-3}K}$ and $\mathrm{6 \times 10^{13} ~cm^{-3}K}$.  To anticipate our results, due to possible contribution of photon-excitation at the line formation, we preferred not to use the line ratio technique to derive the electron density. Instead, these two pressure values are found to provide the better results for the DEM.
Similarly to the QS case, above $\log T = 4.5$ the \ion{S}{} lines are used to constrain the FIP bias. The resulting ratio between the theoretical and observed intensities are listed in Table \ref{tab:ratios}. We see that a pressure of $\mathrm{2 \times 10^{14} ~cm^{-3}K}$ proves to be the best solution. Similarly to the QS results, the DEM does not change when we use the extended line list of Table \ref{table:lines2}.
Both pressures give results in agreement with PAR07, which supports a photospheric composition in prominences.

Looking at the results, we can see  some similarities and  some differences from the results obtained from the  QS data.
In terms of similarities with the QS results,  the newly included \ion{Fe}{iii} ($\log T=4.45$) lines are consistent with the other lines formed at similar temperatures. Also,  the observed \ion{Si}{iv} and \ion{N}{v} lines are too bright with respect to their theoretical value while the \ion{S}{ii} lines and \ion{C}{ii} 1037.018 \AA~ are not bright enough. 

Unlike the QS case, in this prominence the \ion{C}{iii} 977.02 \AA~($\log T=4.7$) observed intensity 
is much higher than predicted (it was excluded from the inversion). 
The  \ion{C}{iii} multiplet  at around 1175 \AA~ as well as those  for \ion{C}{ii} at 1323.9 \AA~ and at  around 1335 \AA~
are marginally  consistent, depending on the pressure chosen. This behavior is expected, given their sensitivity to density 
(see Sections \ref{sec:opacity} and \ref{sec:others} below). The multiplet at around 1335 \AA~ appears to 
be affected by opacity, so  we have excluded it from the DEM inversion, 
even though in some cases the lines intensity can be reproduced by our DEM.

\begin{figure}
\resizebox{\hsize}{!}{\includegraphics[scale=.25]{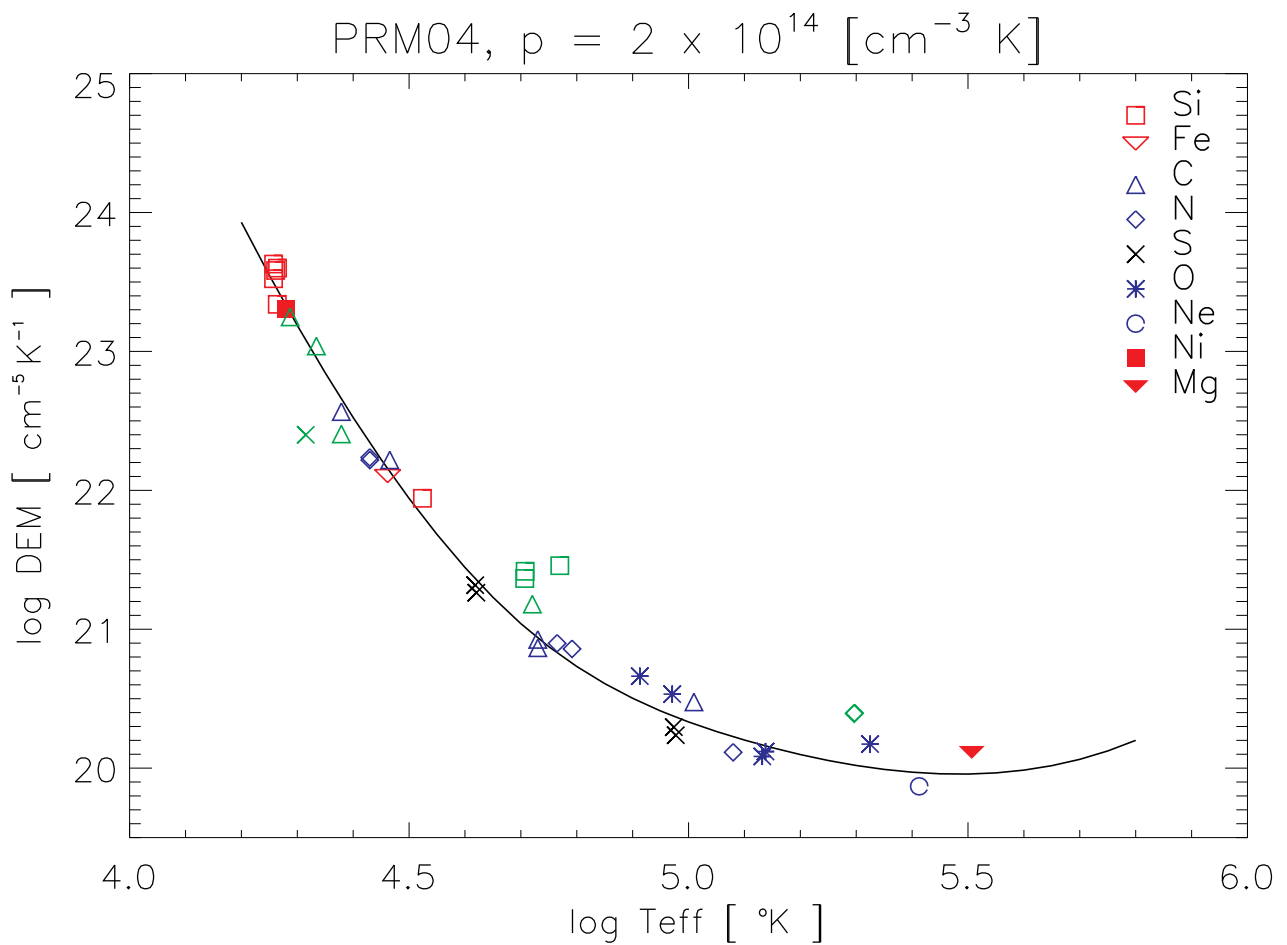}}
\caption{PRM04 DEM assuming a pressure equal to $\mathrm{2\times 10^{14} ~cm^{-3}K}$.}    \label{fig:dem_prmo4}
\end{figure}

We tested the PRM04 data for both pressures used for PRM1 and found that the best solution for the DEM was given by $p=\mathrm{2 \times 10^{14} ~cm^{-3}K}$, which is shown in Figure \ref{fig:dem_prmo4}. 
 Photospheric is also the appropriate composition for this prominence. 
The newly added \ion{Ni}{ii} ($\log T=4.15$) and \ion{Fe}{iii} ($\log T=4.5$) give consistent results. 
And the new \ion{Mg}{v} ($\log T=5.45$) line constrains the FIP bias at the highest temperature. 
Interestingly, we found this dataset to behave in the same way as that from PRM1: the  \ion{S}{ii}, \ion{C}{ii} 1037.018 \AA, \ion{C}{iii} 977.02 \AA, \ion{Si}{iv}, \ion{N}{v} intensities are not consistent with the prediced ones by an amount similar in  sign. 
The  systematic behavior of \ion{S}{ii}, \ion{C}{ii} 1037.018 \AA~ and \ion{C}{iii} 977.02 \AA\ is discussed below. Finally, the use of the extended list of Table \ref{table:lines2}
can also provide information on the temporal variation of the data: lines from the same ion observed 
far away in wavelength (which means a temporal delay greater than one hour) 
provide consistent result (see for instance \ion{Si}{ii}, \ion{C}{ii}, \ion{Si}{iv}).

\subsection{Opacity}
\label{sec:opacity}

It is well known that some of the lower-temperature lines can be affected by opacity; line profiles can become
self-reversed and total intensities decrease. However, these effects are more prominent near the solar limb, and less
in quiet Sun areas on-disk and prominences. 
Our datasets include the \ion{C}{ii} resonant doublet 1037.018 -- 1036.337 \AA~ whose ratio should have a value of two
if the plasma is optically thin. 
This is not the case both in QS (1.2) and prominences (1.2 for PRM1 and 1.4 for PRM04), as already noticed in PAR07. 
Indeed the  brightest 1037.018 \AA~ line is incompatible with our solution, suggesting that some opacity effect may be at work. 
A substantial opacity of these lines in the QS was also reported by other authors \cite[e.g.,][]{chae98, brooks00, giunta15}.

\begin{figure}
\resizebox{\hsize}{!}{\includegraphics[scale=.1]{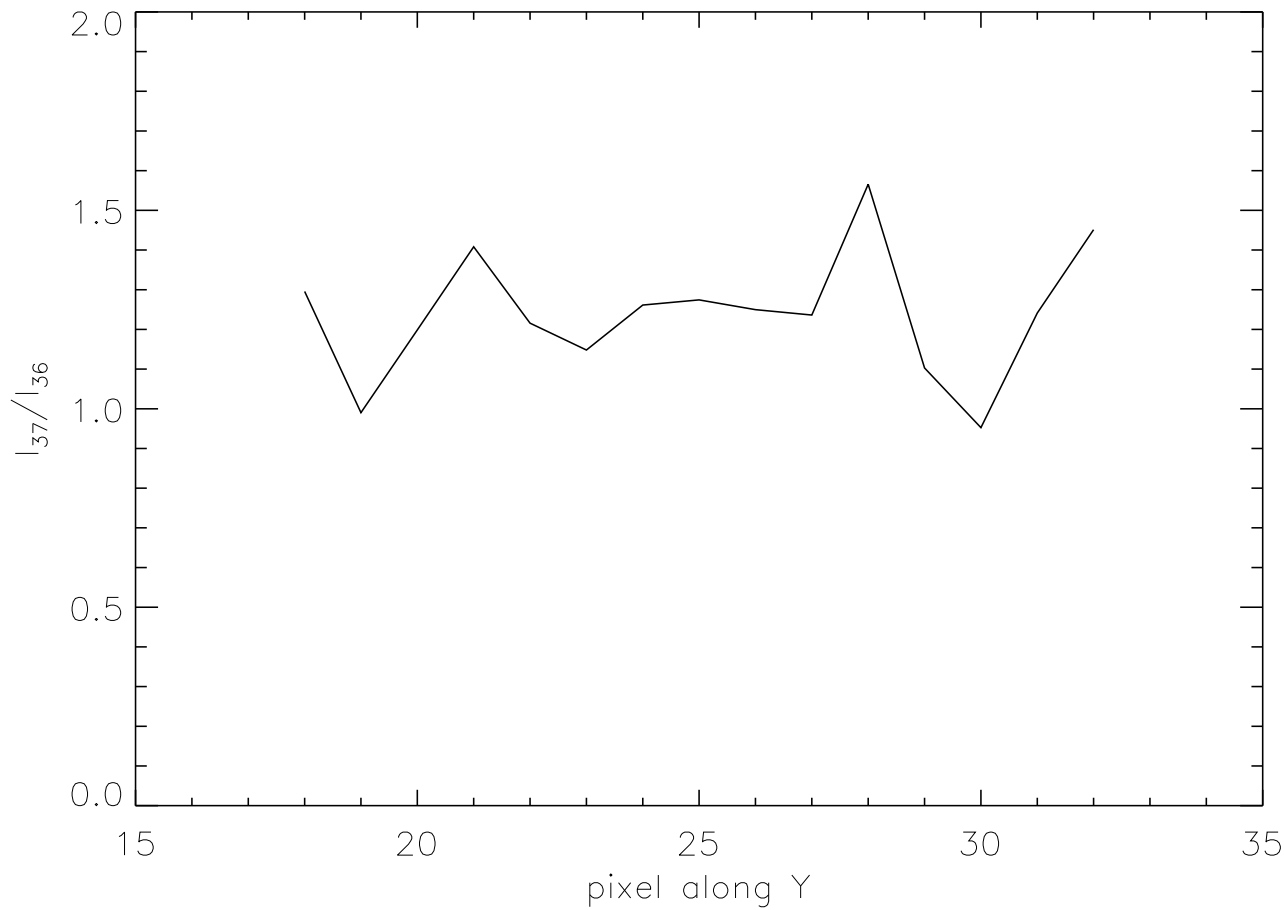}}
\caption{Ratio of the \ion{C}{ii} 1037.018 and 1036.337 \AA~ doublet along the SUMER slit in the pixel range used to build the PRM1 data.}  
\label{fig:c2_prm1}
\end{figure}

We investigated  the behavior of this doublet and found that the line profiles are Gaussian within the prominence area used for the analysis (averaged data within about $15\arcsec \times 1\arcsec$).
The profiles are flat or self-reversed at line center  only close to the limb, which is ouside our range of interest. 
In Figure \ref{fig:c2_prm1} we show that  the intensity ratio within the PRM1 area (using the full resolution of the data)
 is always between 1 and 1.5. 
Even with the medium spatial resolution, we still see variability from pixel to pixel ($\approx$ $1\arcsec$).  
Using the technique of intensity ratios at line center (Equation 9 of \cite{dumont83}) and once the FWHMs are known, one obtains (for the averaged lines profiles) $\tau(1036)$ = 1.2 and $\tau(1037)$ = 2.4 in the QS and  $\tau(1036)$ =1.4 and  $\tau(1037)$ = 2.9 for PRM1,
that is, the opacities are somewhat greater in the prominence than in the QS. 
This result is at odds with the lower intensities in the prominence than in the QS (Table \ref{table:lines}). 
One can suspect that this  has to do with the hypothesis of constant and identical source functions used in Dumont et al.'s technique. We discuss this issue in Section 4.4.

We have also looked at the multiplet at around 1335.5 \AA~ 
(1335.708+11135.663 and 1334.577 \AA);  both lines 
are well reproduced by our DEM. 
However, the CHIANTI database predicts the (1335.708+1135.663)/1334.577 \AA~ ratio to be about two in optically 
thin conditions around the prominence density, while in our case this ratio is 1.45 for the QS, 1.28 for PRM1 and 1.51 for PRM04,
 thus suggesting that some opacity is present. 
 We note that in on-disk QS, chromospheric modeling of these lines gives optically thick conditions \citep[e.g.,][]{ rathore15b}, with a ratio that is variable, depending on the source functions of the lines.
 
In addition, as we will see in the next section, we expect that the intensity of these lines is increased by resonant scattering of the disk radiation, compensating in part for the decrease due to opacity effects. 
As in the case of the  1037.018 \AA~ line, if we apply Dumont et al.'s technique to the PRM1 observations, 
we find that the prominence opacity is larger than in the QS.

We also tested the absorption of other bright lines falling on the \ion{H}{} Lyman continuum. 
Using our resulting DEM for the QS (Figure \ref{fig:demqs}) and prominence PRM1 (Figure \ref{fig:dems_prm1}) and the intensities of \cite{parenti05a}, we found that  
for the \ion{C}{ii} multiplet at around 903.9 \AA~~ (903.6235, 903.9616, 904.1416, 904.4801 \AA) in QS 
the theoretical line intensities are higher by factors between 2.2 and 4.6 with respect to the observed values.
 In the case of the PRM1 we found even higher values (between 4.8 and 10). 
This indicates that additional absorption  comes from the Lyman continuum. 
 A similar behavior is found for the multiplet of \ion{O}{ii} and \ion{O}{iii} at around 833--835 \AA. 
For instance, the $\mathrm{I_{th}}/\mathrm{I_{ob}}$ for \ion{O}{ii} 834.466 \AA~  is 4.6 for PRM1 and 3.9 for the QS. 
For the \ion{O}{iii} 833.715+ 833.749 \AA~ we have 4.1 in PRM1 and 3.6 in the QS, supporting our choice of excluding
these lines from the DEM analysis.

\subsection{Other contributions to the line formation}
\label{sec:others}




Lines that produce strong disk illumination may have non negligible radiative excitation within prominences.
Our  results in terms of abundance diagnostics would be consequently affected. 
A detailed assessment of this effect is not trivial, as it would require knowledge of how the density and 
temperatures are distributed within the PCTR, how many PCTRs are along the line of sight and at what distance from the 
disk, as well as how much disk radiation reaches the prominence. 
The local contribution due to resonant excitation is in fact dependent on the local 
densities and temperatures. For lower densities, the  contribution due to the resonant excitation would relatively increase. 

We start by considering one of the strongest EUV lines, the \ion{C}{iii} 977.020 \AA.
\cite{jordan01} found the \ion{C}{iii} 977.020 \AA~  optically thick in some areas of the QS, 
while \cite{doschek04} found this line optically thin at disk center. 
In our analysis we have shown that the  line is always too bright for our DEM solutions, which 
suggests that  photo-excitation could be increasing its intensity.
\cite{jejcic17}, for instance, found the radiative excitation to be important under 
their low density (much lower than our case) erupting prominence. 

\begin{figure}
\resizebox{\hsize}{!}{\includegraphics[scale=.5]{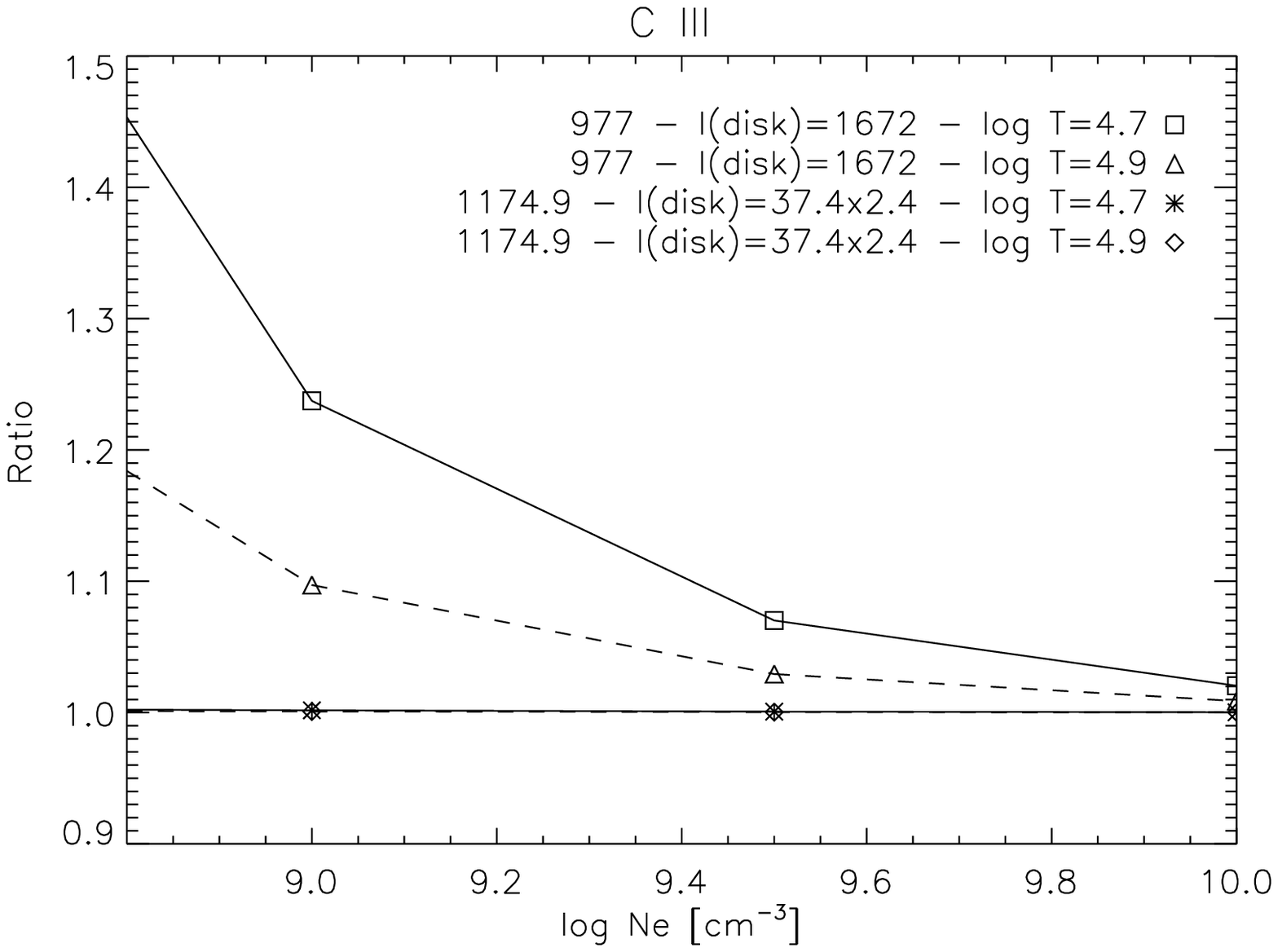}}
\resizebox{\hsize}{!}{\includegraphics[scale=.5]{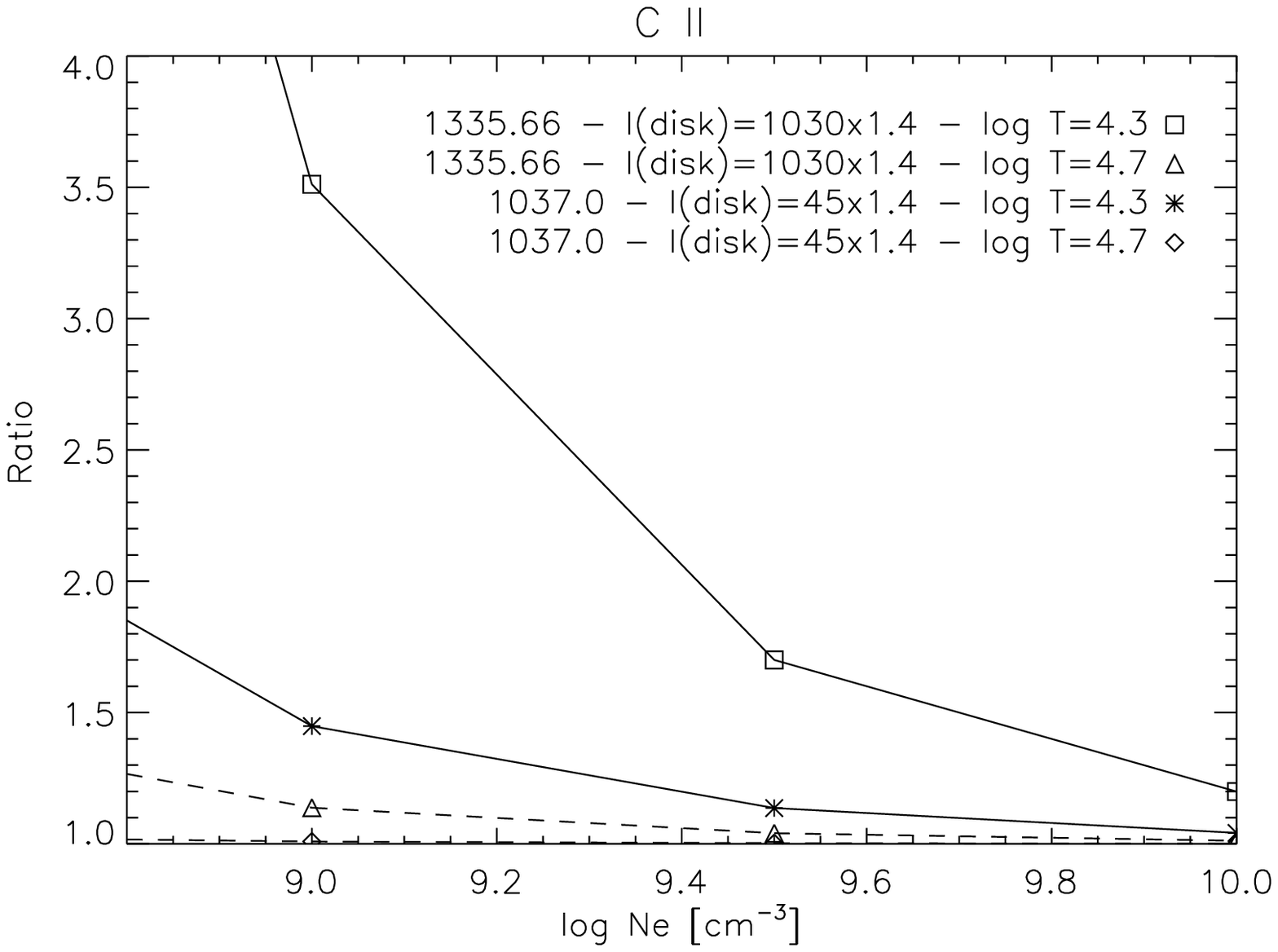}}
\caption{Top: Ratio of the total emissivity (including photo-excitation) vs. the values without
photo-excitation, for a range of densities and two values of the temperature, for two  \ion{C}{iii} lines.
Bottom: Same as Top, but for two  \ion{C}{ii} lines, at 1037.01 \AA~ and 1335.66~\AA.
} 
\label{fig:c_phot}
\end{figure}

We have assumed a distance of the prominence from the solar chromosphere of 26 Mm. 
For the disk radiance, we have taken the averaged on-disk quiet Sun value obtained by 
SUMER radiance measurements \citep{wilhelm98c}, 1672 ergs cm$^{-2}$ sr$^{-1}$ s$^{-1}$. 
We note that this is significantly higher, by a factor of 2.4, than the 
radiance near disk center. 
From the DEM modeling, the effective formation temperature
of \ion{C}{iii} is at around log $T$[K]=4.7 using the OPEN-ADAS rates, compared to the peak of the 
emissivity at around  log $T$=4.9. Figure \ref{fig:c_phot} (top) shows the increase in the 
emissivity of the \ion{C}{iii} 977.02~\AA~ line due to photo-excitation, for a range of densities and
these two temperatures. We note that a pressure of $\mathrm{6\times 10^{13} ~cm^{-3}K}$
would correspond to a density of $ \mathrm{10^{9}~cm^{-3}}$ at  log $T$[K]=4.7, and an increase of emissivity of  about 25\% . Increasing the  pressure, the collisional excitation will be even more important, 
making the radiative excitation probably irrelevant.

On the other hand, the other  \ion{C}{iii} lines of the multiplet, being very weak,
are not affected by photo-excitation. The same figure (\ref{fig:c_phot} top) shows one of them, the line at 
1174.9~\AA, for which we have assumed an averaged disk radiance 2.4 times a 
disk center value of 37.4 ergs cm$^{-2}$ sr$^{-1}$ s$^{-1}$.

We have carried out the same estimate for  \ion{C}{ii} lines, shown in 
Figure \ref{fig:c_phot} (bottom). 
We have considered the strongest transition of the doublet, 
the \ion{C}{ii} 1335.66  \AA\ line, which shows a significant increase. The disk radiance 
is somewhat uncertain. We have used the total disk radiance of the doublet,
1030 ergs cm$^{-2}$ sr$^{-1}$ s$^{-1}$, and increased it by a factor of 1.4 to take into 
account limb-brightening effects. 
We note that a pressure of $\mathrm{6\times 10^{13} ~cm^{-3}K}$
would correspond to a density of $\mathrm{10^{9.5}~cm^{-3}}$ at  log $T$[K]=4.3, and an increase of 
about 70\%. There is a strong temperature sensitivity though, so it is more likely
that the increase is less than this value. 

On the other hand, following this calculation, one of the lines of the multiplet at shorter wavelengths, the 
\ion{C}{ii} 1037.0 \AA, is not affected so much. We have assumed for the disk
radiance 45 ergs cm$^{-2}$ sr$^{-1}$ s$^{-1}$, and increased it by a factor of 1.4 to take into 
account limb-brightening effects. The increase is on the order of 10--20\% at most.

We see that for this line, the QS opacity is lower than the prominence value. This surprising result can be explained when one takes into account that the QS line results from electron collisions while the prominence line may be partly formed through resonance scattering of the incident chromospheric \ion{C}{ii} radiation. Consequently the source functions of the two lines of the doublet are not equal and tend to be in the ratio of the two incident line intensities. 
Our simulation of a weak contribution of the resonant scattering for this line may  suffer from the uncertainty of the disk illumination. Our only reference \citep{wilhelm98c} for the solar irradiance refers to solar minimum data.  In \cite{parenti04} we estimate that during our observations about 20$\%$ of the disk was occupied by active regions, even though no single active region was close by. 
 
In summary, considering the uncertainties in the estimates of the 
 photo-excitation effects, it is better to use the weaker lines for the DEM analysis. 
The use of  optically thin lines from the same ion such as the multiplet at \ion{C}{ii} 1323.8 \AA~ 
helps better constrain the DEM at such temperatures.

The \ion{S}{ii} 1253.79--1250.58 \AA~ are also not consistent with our DEMs, being less intense.
 At the same time, as they also are bright lines, it is possible that their radiative component is not negligible. 
In fact, the ratio of the intensities is 2.36 in the QS and 3.24 in the PRM1 with respect to the theoretical value of two (which corresponds to the collision-only situation).

In our analysis we show how some of the Li-like and Na-like iso-electronic sequence ions produce lines with intensities much larger than what is predicted by our DEMs. This is also the case for the doublet \ion{Si}{iv} 1393.757 \AA~ and 1402.772 \AA~  from PRM04. We would like to point out also that the observed intensities are not in a ratio two as expected, but greater. This can be an indication of a radiative contribution to the line. This doublet has ratio 2 in QS \citep{chae98} but can be higher in brighter areas such as  AR \citep{gontikakis18}.

Finally, we also find \ion{O}{v} 1218.347 \AA~ to be too bright for our DEMs. This line lies in the \ion{H}{i} Lyman $\alpha$ red wing and it is possible that the total intensity receives a contribution from the scattering of this very bright and extended wing.

\section{Summary and conclusions}
\label{sec:disc}

The main aim of this work was to establish the elemental composition and FIP bias in quiescent prominences. We have analyzed two prominences that reveal consistent results, both supporting photospheric composition (and thus no FIP bias). We used the DEM technique, which requires us to invert a set of spectral line intensities whose formation temperatures span  the relevant range. 
 
In our previous works on these prominences, for each ion we used as many spectral lines as possible in order to provide the average DEM value at that ion temperature. 
 Here instead we limited our line list  to the most reliable ones at a given temperature. This novelty allows to minimize the uncertainties in the inferred abundances. Additionally, we restrained the analysis to the low transition region, in order to limit the effect in our results of possible FIP bias variation with temperature. 
 Improvement with respect to our previous analysis are also the use of density-dependent ionization fraction data and the inclusion of additional spectral lines (Table \ref{table:lines2}) and ions (\ion{Fe}{iii}, \ion{Ni}{ii}, \ion{Mg}{v}). 
 With regard to our previous analysis in \cite{parenti07} and \cite{gunar11}, these additional ions impose a stronger constraint to the low FIP bias. 
Beside the different approach in the diagnostics of this work,  we confirm our previous finding: a photospheric composition in prominences and quiet Sun transition region.

To explain some inconsistencies in the results, we also investigated the possible presence of opacity or/and resonant scattering processes that contribute to the line formation and which are not compatible with the DEM diagnostic technique (unless corrections to the line intensity are made). We show that in prominences and the quiet Sun transition region the bright doublets \ion{C}{ii} and \ion{S}{ii} are affected mostly by opacity (as well as the lines lying on the \ion{H}{i} continuum), and possibly also by resonant scattering from the disk illumination. 
The  \ion{C}{iii} 977.020 has  a non-negligible scattering component in prominences. Nevertheless, we find that \ion{C}{ii} 1335.663 and 1334.577 \AA~ intensities agree with the optically thin 1323.862 \AA~ multiplet for our DEM solution. Thus, it is possible that opacity and radiative scattering compensate for each other. Additional modeling would be needed to investigate further these lines.  For the reasons mentioned here we excluded all these lines from the DEM inversion and we warn the reader against use of these lines with this technique. 

We note that some inconsistencies found in our work may be the result of the uncertainty in the ionization equilibrium calculation of the ions. For instance, the  \ion{C}{ii} ratio 1036.337/1334.6 \AA~ is strongly temperature-dependent (but independent of the ionization equilibrium), and it can be used to get an average temperature of formation of the ion which, in our case, gives about $\mathrm{\log T = 4.2}$. This value is lower than the one  predicted with CHIANTI and much closer to the value predicted by ADAS (see Figure \ref{fig:ioneq}). It is possible that the ion is formed at even lower temperatures. The results of this work are a contribution to the understanding of the origin of the prominence plasma: models that predict coronal condensation process (from plasma with coronal composition) as the filling process of the filament channel, have to be excluded or further improved to be compatible with  our results.\\


The analysis of the upcoming data from the ESA-NASA Solar Orbiter mission will benefit from our results, as we provide several guidelines for the data interpretation. The SPICE spectrometer will cover, co-temporally, the spectral ranges  704--790 \AA~ and 972--1005 \AA~ (first order). It will  observe several lines that have been used in this work, and it will cover most of the ions that we listed (see Tables \ref{table:lines} and \ref{table:lines2}). Among these there are the  \ion{C}{ii} 1031--37 \AA~ and \ion{C}{iii} 977 \AA~ lines, which we have shown to have a complex line formation. These lines will be observed frequently, as they are among the brightest lines in the UV prominences and QS spectra. We then 
have provided a guide to the interpretation of their radiance, by clearly showing the differences in their properties between the QS and prominence plasmas, and the need for taking into account  multiple processes in the line formation. 

We also recommend that future observations should be accompanied by the development of accurate atomic models. The existing CHIANTI and ADAS databases are already an excellent reference for this. However, improvements are foreseen, and we identified a few aspects that can be further investigated and for which the interpretation of the SPICE data, and all the future missions data, will benefit. It has already been shown that the analyses made using less accurate atomic models, or using diagnostic techniques that are correct only under simplified plasma atmospheres, 
resulted for instance in a too strong coronal FIP bias. This would have provided  erroneous inputs to the problem of connectivity between the Sun and the heliopshere. 

We also highlight those lines used in this work for which we are confident in the optically thin approximation (in both QS and prominences), dominant collisionally excited   level population process, with a generally correct atomic models (when all the "ingredients" we have discussed are taken into account). These include several lines also to be observed by SPICE. Even if we have not exploited the full SPICE possibilities here (it was not the purpose of the work), with our line list we can already sample in temperature the chromosphere and low transition region plasma quite accurately and cover both low and high FIP groups.
The Solar Orbiter  in-situ suite SWA, with the instrument HIS, will measure heavy ions abundance such as \ion{Fe}{}, \ion{Si}{}, \ion{Mg}{}, \ion{C}{}, \ion{N,}{} and \ion{O}{} allowing, together with SPICE, to map the FIP bias back on the Sun.

\begin{acknowledgements}

SP acknowledges the funding by CNES through the MEDOC data and operations center.
This work used data provided by the MEDOC data and operations center (CNES / CNRS / Univ. Paris-Sud), http://medoc.ias.u-psud.fr/.
GDZ  acknowledges support by  STFC (UK) via the
consolidated grant of the DAMTP atomic astrophysics group at the University of
Cambridge.

This work was carried out as part of the research of the
 ISSI international team n. 418, {\it Linking the Sun to the Heliosphere using Composition Data and Modelling }. 
CHIANTI is a collaborative project involving George Mason University, the University of Michigan (USA) and the University of Cambridge (UK). 
SUMER is financially supported by DLR, CNES, NASA, and the ESA PRODEX program
(Swiss contribution). SOHO is a mission of international cooperation
between ESA and NASA.
The data for this work were taken during two MEDOC campaigns at IAS, France.  
We acknowledge the use of the OPEN-ADAS database, maintained by the 
 University of Strathclyde.
SP acknowledge John Leibacher and the anonymous referee for their suggestions that have improved the clarity of the paper.

\end{acknowledgements}

\bibliographystyle{aa} 

\begin{appendix}
\label{appendix1}
\onecolumn    
\section{Table with results from the DEMs inversion}

\begin{table*}[hb]
\caption{Ratios (R) of predicted vs. observed radiances of the selected lines for the Various tests. After the line identification, we list the $T_{\rm eff}$, the results for, respectively, QS, PRM1 and PRM04 (the assumed pressures are in $\mathrm{cm^{-3}K}$).  The '*' means that it was not included in the DEM inversion but the radiances are calculated using the resulted DEM. }             
\label{tab:ratios}      
\centering          
\begin{tabular}{l l c c l l @{\hspace{-0.3cm}}l @{\hspace{0.5cm}}l @{\hspace{-0.3cm}}l @{\hspace{0.5cm}}l @{\hspace{-0.3cm}}l}     
\hline\hline       
Ion & $\lambda_{th} [\AA]$ & log $T_{\rm eff}$[K]& QS & & PRM1 & & PRM1 & & PRM04 &\\ 
    &                       &                     &   & & $\mathrm{2\times 10^{14}}$& & $\mathrm{6\times 10^{13}}$ && $\mathrm{2\times 10^{14}}$ &\\   
\hline                    
   Si II  & 1190.416 & 4.26 & 0.62 & & 1.01 && 0.92 & &0.69 &\\
   Si II  & 1197.3955& 4.26 & 1.11 &&      &&       &&1.27 &\\ 
   Si II  & 1264.738 & 4.26 &      &&      &&       && 0.76 &\\ 
   Si II  & 1304.370 & 4.26 &      &&      &&       && 0.73 &  \\ 
   Si II  & 1309.276 & 4.26 &      &&      &&       && 0.93 &\\
   Ni II  & 1317.210 & 4.28 &       &&     & &      && 1.05 &\\
   S II   & 1253.811 & 4.30 & 1.51 &*& 4.3 &*& 4.2 &* & 4.80&* \\
   S II   & 1250.585 & 4.30 & 1.74&*& 6.7 &*& 6.60 &* &   &\\
   N II   & 1085.701 & 4.41 & 1.02 && 0.83 && 0.77 &&1.28&\\
   N II   & 1083.990 & 4.41 & 0.92 && 0.79 && 0.81 &&1.33&\\   
   Fe III & 1017.254 & 4.45 & 0.87 && 0.89 && 0.66 & &1.26&\\
   Fe III & 1128.740 & 4.37  &      && 0.88 && 0.77 &&   &\\  
   C II   & 1335.663 (2)& 4.27 & 1.36&* & 0.92 &*& 0.83 &* & 1.07 &* \\
   C II   & 1334.577 & 4.32 & 1.00&* & 0.78&* & 0.71 &*& 0.81 &* \\
   C II   & 1036.337 & 4.36 & 1.27 && 0.95 && 0.98 && 1.24 &\\
   C II   & 1037.018 & 4.36 & 2.07&*& 1.58 &*& 1.64 &*& 1.79&* \\ 
   C II   & 1323.862 (4)& 4.43 & 0.87 && 0.76& & 0.16 &* & 0.82 & \\
   Si III & 1206.502 & 4.42 & 0.71 && 1.31& & 1.45 & &   &\\    
   Si III & 1109.943 & 4.48  &      && 0.85 &&0.93  &  & &\\ 
   Si III & 1113.25  & 4.48 & 0.64  && 0.80 &&   0.92 &&  0.75&\\ 
   S III  & 1200.959 & 4.58 & 1.01 && 1.54 && 1.75 &&1.11 &\\
   S III  & 1015.496 & 4.61 & 0.92 & &1.23 && 1.42 &&    &\\
   S III  & 1194.047 & 4.62 &       &&    &&      &&  1.23 & \\ 
   C III  & 977.020  & 4.66 & 1.28  && 0.46 &*&  0.55&*& 0.61&*\\
   C III  & 1174.933 & 4.69 & 0.88  && 0.77 &&0.60 &*  & 1.18 & \\
   C III  & 1175.7111& 4.69 &       && 0.70 && 0.54 &* &1.03&\\
   Si IV  & 1128.34  & 4.71 & 0.11&*& 0.13 &*&  0.15 &* & 0.23&*\\
   Si IV  & 1393.76  & 4.71 &      &&      &&       &&0.39&* \\
   Si IV  & 1402.77  & 4.71  &      &&     & &       &&0.45&*\\  
   N III  & 989.799  & 4.76 & 1.07 && 0.73 && 0.78 &&0.85 &\\
   N III  & 991.577  & 4.76 & 0.97  && 0.66 && 0.66 & & 0.79&\\
   O III  & 703.850  & 4.91 &       &&      & &     && 0.66  & \\
   O III  & 525.794  & 4.93 & 1.21  &&      &&      & &  0.70 &\\
   S IV   & 1072.974 & 4.93 & 0.74 && 1.30 && 1.27 & & 1.20 &\\
   S IV   & 1062.664 & 4.94 & 0.97 && 1.19 && 1.38 &&1.35 &\\
   O IV   & 1399.776  & 5.13 &      &&      &&      & & 1.21 &\\
   O IV   & 1401.163  & 5.14 &      &&      &&      & & 1.10 &\\
   O IV   & 555.264  & 5.18 & 0.92 &&   &&       & &    &\\
   C IV   & 1548.187 & 5.01  &      &&      &&      & & 0.69 &\\
   N IV   & 765.147  & 5.08 &      &&      &&       && 1.29 &\\ 
   N IV   & 955.334  & 5.24 & 0.99  && 0.89 && 0.89 & &  &\\
   O V    & 1218.347 & 5.32 & 0.47 && 0.62 && 0.62 && 0.68 &\\
   N V    & 1238.821 & 5.40  & 0.62 &* & 0.32 &* & 0.32 &* & 0.42&* \\
   N V    & 1242.804 & 5.40 & 0.61 &* &  0.34 &*&  0.34 &* & 0.43&* \\
   Ne V   & 1145.582 & 5.42  & 1.19 & & 1.25 &&  1.42 && 1.25  &\\
   Ne V   &  572.113 & 5.45 & 0.82 && 0.79 && 0.91 & &  &\\
   Mg V   & 1324.433 & 5.51  &      &  &   &     &  && 0.78& \\
   O VI   & 1031.912 & 5.63 & 1.01 &&      &&      & &  &\\
   O VI   & 1037.614 & 5.63 & 1.07 &&      &&      & & &\\
\hline                  
\end{tabular}
\end{table*}

\end{appendix}
\twocolumn

\end{document}